# Photoluminescence-Based Gas Sensing with $MoS_2$ Monolayers

**GIA QUYET NGO,[1,2]* CHANAPROM CHOLSUK,[1,3,4] SEBASTIAN THIELE,[5] ZIYANG GAN,[6] ANTONY GEORGE,[6] JÖRG PEZOLDT,[5] ANDREY TURCHANIN,[6] TOBIAS VOGL,[1,3,4], AND FALK EILENBERGER[1,2,7]**

[1]*Institute of Applied Physics, Abbe Center of Photonics, Friedrich Schiller University Jena, Albert-Einstein-Str. 15, 07745 Jena, Germany*

[2]*Fraunhofer-Institute for Applied Optics and Precision Engineering IOF, Albert-Einstein-Str. 7, 07745 Jena, Germany*

[3]*Department of Computer Engineering, TUM School of Computation, Information and Technology, Technical University of Munich, Theresienstr. 90, 80333 Munich, Germany*

[4]*Munich Center for Quantum Science and Technology (MCQST), Schellingstr. 4, 80799 Munich, Germany*

[5]*FG Nanotechnologie, Institute of Micro- and Nanoelectronics and Institute of Micro- and Nanotechnologies MacroNano and Institute of Materials Science and Engineering, Technical University Ilmenau, Gustav-Kirchhoff-Str. 7, 98693 Ilmenau, Germany*

[6]*Institute of Physical Chemistry, Abbe Center of Photonics, Friedrich Schiller University Jena, Lessingstraße 10, 07743 Jena, Germany*

[7]*Max Planck School of Photonics, Germany*

**quyet.ngo@uni-jena.de*

**Abstract:** Two-dimensional transition metal dichalcogenides (TMDs) are highly appealing for gas sensors, lab-on-a-chip devices and biosensing applications because of their strong light-matter interaction and high surface-to-volume ratio. The ability to grow these van der Waals materials on different substrates and waveguide geometries opens a horizon toward scalable on-chip photonic nanodevices. Here we report on a versatile technique for remote optical gas sensing using two-dimensional TMDs. The adsorption of the gas molecules on the monolayer surface provides a gateway for gas sensing based on charge-transfer-induced photoluminescence variation. For gases that are weakly adsorbed on the surface of monolayer TMDs, purging the monolayers' surface by an inert gas like $N_2$ can desorb gases from the monolayers at room temperature. We demonstrate CO, NO and $NO_2$ detection by monitoring photoluminescence from semiconducting $MoS_2$ monolayers grown on $SiO_2$/Si chips at a level of 10 ppm with fast response time. Observations are supported by our density functional theory calculations, which predict a significant interaction between these gases and $MoS_2$ monolayers. These findings may lead to advances in remote sensing, surface-sensitive bioanalytics and lab-on-a-chip sensors.

## 1. Introduction

Two-dimensional (2D) materials are emerging as potential candidates for gas sensing due to their unique physical and chemical properties and their high surface-to-volume ratio. Nanostructured materials are highly appealing for gas sensing at low operating temperatures [1], and graphene has shown the capability of detecting single $NO_2$ molecules at room temperature [2]. Besides graphene, transition metal dichalcogenides (TMDs) are highly attractive for photonics and optoelectronics because of the nature of direct-gap semiconductors at monolayer limits [3]. The strong photoluminescence (PL) of 2D TMDs paves a novel way toward optical gas sensing. In addition, the conformal growth of semiconducting monolayers on different geometries and materials is unique allowing for integrated 2D functionalized circuits and 2D materials-based sensors [4]. 2D TMDs have revealed their significant potential for gas sensing due to the large surface-to-volume ratio and ability to engineer surface activities by defects [5,6]. When TMD crystals are exposed to a target sensing medium, their change of optical and electronic properties is expected to be larger compared to their bulk counterparts.

There are several processes involved in gas sensing, for example, the physisorption and chemisorption of gas molecules on the sensing material's surface [4−11]. The adsorption of gas molecules can alter the optical properties of 2D TMDs, for example, via modification of the dielectric constant of monolayer $MoS_2$ by $O_2$, $NO_2$, and NO molecules [12]. The charge transfer process depends on the doping role of the gas to the sensing material [8,9] and originates from the physisorption of gas molecules on the sensing material's surface [13,14]. The charge transfer can accumulate or deplete charge carriers in TMDs [15,16]. Because the PL characteristics in monolayer TMDs are driven by excitons, any change in the concentration of charge carriers will induce an enhancement or reduction in the PL intensity and/or a shift in the PL emission peak. The PL signal of exfoliated monolayer $MoS_2$ was shown to increase 100 times when exposed to $O_2$ and $H_2O$ in a vacuum chamber, where $O_2$ and $H_2O$ induce molecular gating [14]. Hence, the PL spectrum and its features, for example, peak shift and intensity change, can be used for gas sensing. Therefore, monolayer TMDs are a promising candidate for PL-based sensing.

There are many methods to detect the target gases, for example, via absorption spectroscopy [17], photoacoustic spectroscopy [18], photothermal spectroscopy [19], and Raman spectroscopy [20]. Recently, Pasupuleti *et al.* have reported a highly sensitive $H_2S$ gas sensor using CuO@$V_2C$ MXene van der Waals heterostructure [21] while Pham *et al.* have reported an enhancement of $NO_2$ gas sensing using $TiO_2$@Pd nanostructures [22]. In this work, we investigate and demonstrate sensitive but non-selective optical remote gas sensing with a fast response time using PL spectroscopy. We grow monolayer $MoS_2$ directly on $SiO_2$ substrates by a chemical vapor deposition (CVD) [23−27] and use them as active sensing elements. We exploit the PL of monolayer $MoS_2$ and observe its variation as a function of gas exposure. The monolayer $MoS_2$ has long-term PL stability in contrast to the photobleaching of organic fluorophores [28,29]. The samples employ the charge transfer mechanism between monolayers and sensing medium to detect the target. Together with density functional theory (DFT) calculation, we characterized the response of monolayers on different gases. The capability of $MoS_2$ in optical gas sensing is experimentally proved with a detectivity level of 10 ppm. The measurements exhibit potential in optical gas sensing of certain toxic gases. Our findings can also apply to more complex photonic platforms, for example, 2D materials integrated on waveguides [23,24,30−32] and can open up to waveguide-based sensors.

## 2. Results and Discussion

### Density Functional Theory

DFT is a powerful tool for calculating the adsorption energy between gas molecules and TMD monolayers [15]. In this section, we aim to theoretically investigate the possibility of using $MoS_2$ as gas sensors. The influence of gas molecules on the band structure of the monolayer TMDs is first investigated via the DFT calculations using the Vienna Ab initio Simulation Package (VASP). See the Method Section for more computational details. Figure 1 displays the electronic band structures of the pristine $MoS_2$ and $MoS_2$ monolayer in the interaction with CO, NO, and $NO_2$ molecules with spin polarization included. We found that the direct gap of pristine $MoS_2$ remains unaffected by the gas molecule adsorption. While NO and $NO_2$ induce some states to the electronic structures of the monolayer, CO remains unchanged from the pristine electronic structures. This indicates that $MoS_2$ is likely most sensitive to NO and $NO_2$. For NO, one occupied and two extra unoccupied states are formed near the conduction band, referring to the n-type doping effect. On the other hand, for $NO_2$, an extra spin-down unoccupied state is localized near the valence band, implying p-type doping.

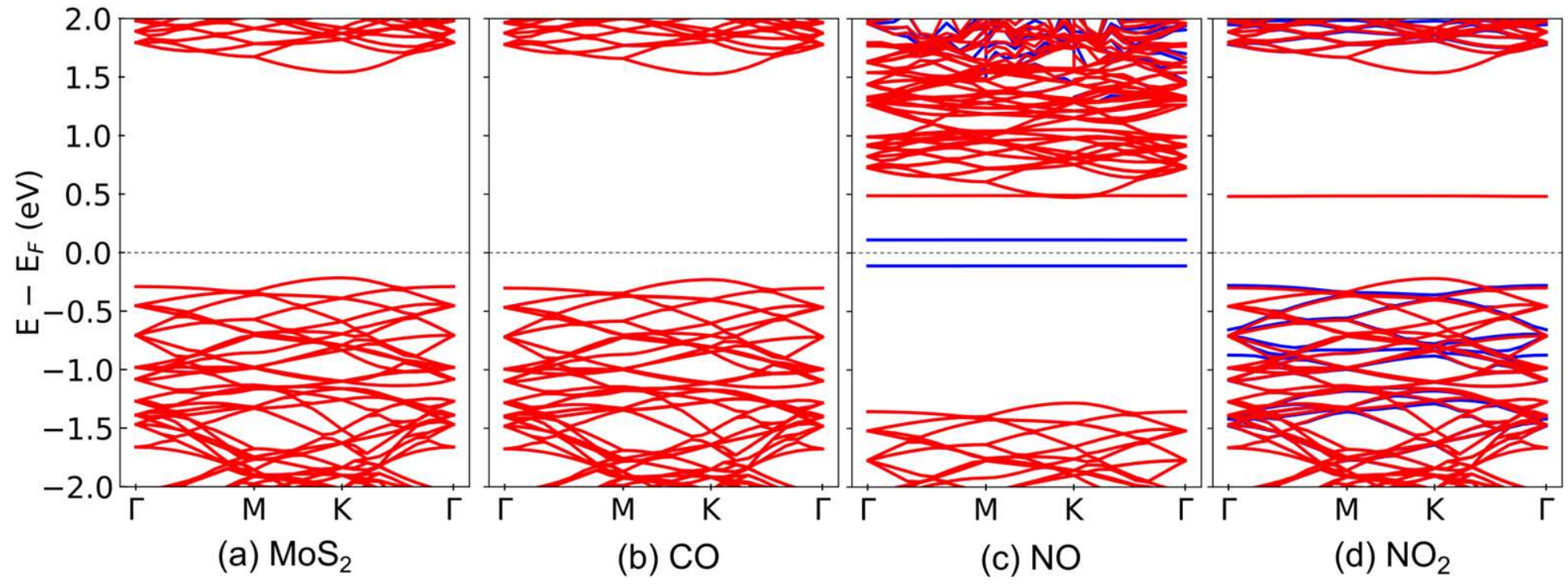

**Fig 1.** Electronic band structures of $MoS_2$ monolayer in the interaction with the target gases. (a) Pristine $MoS_2$. (b) With CO. (c) With NO. (d) With $NO_2$. Calculations for spin-up and spin-down configurations are in blue and red, respectively.

For the quantitative aspect of gas adsorption, the adsorption energy ($E_{\mathrm{ad}}$) needs to be determined. This can be calculated by Equation 1.

$$E_{\mathrm{ad}} = E_{\mathrm{2D+g}} \quad - E_{\mathrm{2D}} - E_{\mathrm{gas}}, \qquad (1)$$

where $E_{\mathrm{2D+gas}}$ is the total energy of the monolayer absorbed with the gas molecule, $E_{\mathrm{2D}}$ is the total energy of the monolayer, and $E_{\mathrm{gas}}$ is the total energy of the isolated gas molecule.

**Table 1.** Calculated adsorption energy for target gases with $MoS_2$ monolayer

| **Gases** | **Adsorption energy (eV)** |
|---|---|
| CO | 0.004 |
| NO | -0.019 |
| $NO_2$ | -0.321 |

Turning to gain more insights into how the charge transfers between monolayer and gas, we computed the charge density difference ($\Delta\rho$) obtained by

$$\Delta\rho = \rho_{\mathrm{2D+gas}} - \rho_{\mathrm{2D}} - \rho_{\mathrm{gas}}, \qquad (2)$$

where $\rho_{\mathrm{2D+gas}}$, $\rho_{\mathrm{2D}}$, and $\rho_{\mathrm{gas}}$ are the charge densities of the gas-adsorbed monolayer, the bare monolayer, and the isolated gas molecule, respectively. The positive (negative) charge density difference implies charge accumulation (depletion).

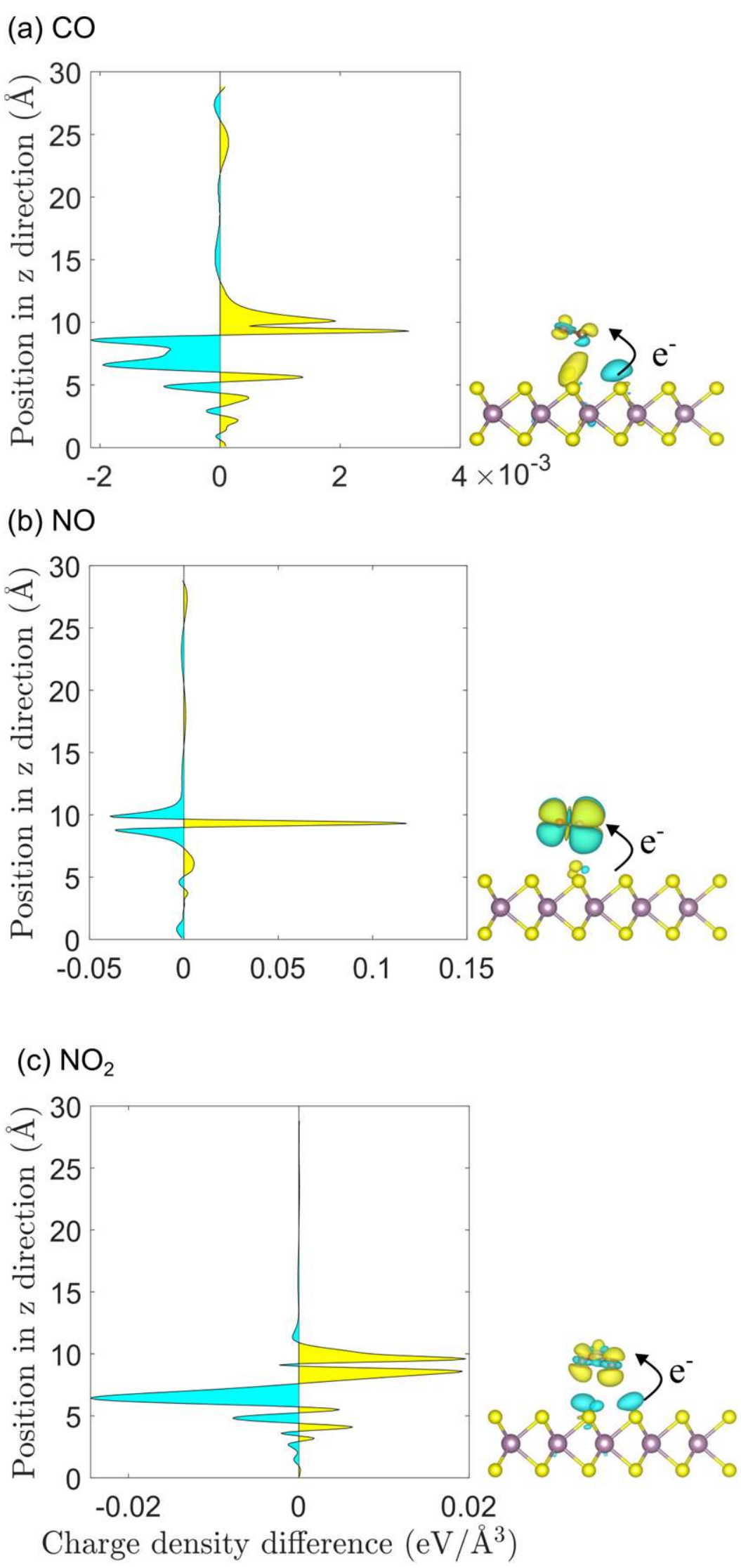

**Fig 2.** Charge density difference ($\Delta\rho$) of $MoS_2$ absorbed by (a) CO, (b) NO, (c) $NO_2$ where the positive charge difference or yellow isosurface represent charge accumulation whereas the negative one or cyan isosurface represent charge depletion. The y-axis signifies the charge density difference at each position in the z-direction corresponding to the atomic position. The yellow and purple atoms illustrate sulfur and molybdenum, respectively.

In principle, the adsorption affects the PL intensity observed in the experiment, depending on the type of monolayers. The negative (positive) adsorption energy implies the gas is likely (unlikely) absorbed on the monolayer. Moreover, the specific value is important for the required temperature of operation. Values much larger than $k_BT$ (25 meV) will lead to strong adsorption and require heating to reset the sensor. Table 1 points out that NO and $NO_2$ gases yield negative adsorption energy for $MoS_2$, reflecting that these gases can be adsorbed. Finally, CO yields a value very close to zero but may need further investigation to confirm the interaction with $MoS_2$ as the calculated value of adsorption energy is less evident than others. As shown in a previous report [15], this may also be due to the functionals used.

Instead of calculating the total charge within an atomic volume, the so-called Bader charge like other DFT calculations [15,16,33], we computed the charge density difference at each specific distance from the monolayer to the gas. This way, we can visualize and capture the charge transfer process between the two systems. As shown in Figure 2, CO, NO, and $NO_2$ molecules predominantly act as charge acceptors when interacting with the $MoS_2$ monolayer. This is evidenced by the presence of a yellow peak with a positive charge density difference in the gas region, indicating charge accumulation. Meanwhile, charge depletion at the $MoS_2$ surface is observed as a cyan peak, confirming that the monolayer donates charge to the gas molecules. These findings are consistent with previous Bader charge analyses [16,33], ensuring the role of $MoS_2$ as a charge donor in these gas adsorption interactions.

In summary, although DFT calculations are based on 0 K, the results can still guide our experiment on the possible behaviors of gas adsorbed on pristine monolayers once pumped into the chamber. Comparison between electronic structures with and without gas indicates the changes in electronic states due to the impurity's state occupation. Together with adsorption energy, the calculation pinpoints that NO, and $NO_2$ are likely adsorbed on $MoS_2$ monolayer, but CO seems vague. All three gases seem to accept charges from $MoS_2$. Our calculation can, therefore, suggest a clue on the PL intensity through the response of each gas. Note that the current DFT simulation focuses only on the adsorption process on the pristine monolayer. However, the $MoS_2$ monolayer itself can contain some intrinsic defects such as vacancies and grain boundaries, which can influence the adsorption energies and charge transfer. This calculation demands lots of computational power and is, hence, beyond the scope of this work.

**Photoluminescence-Based Gas Sensing**

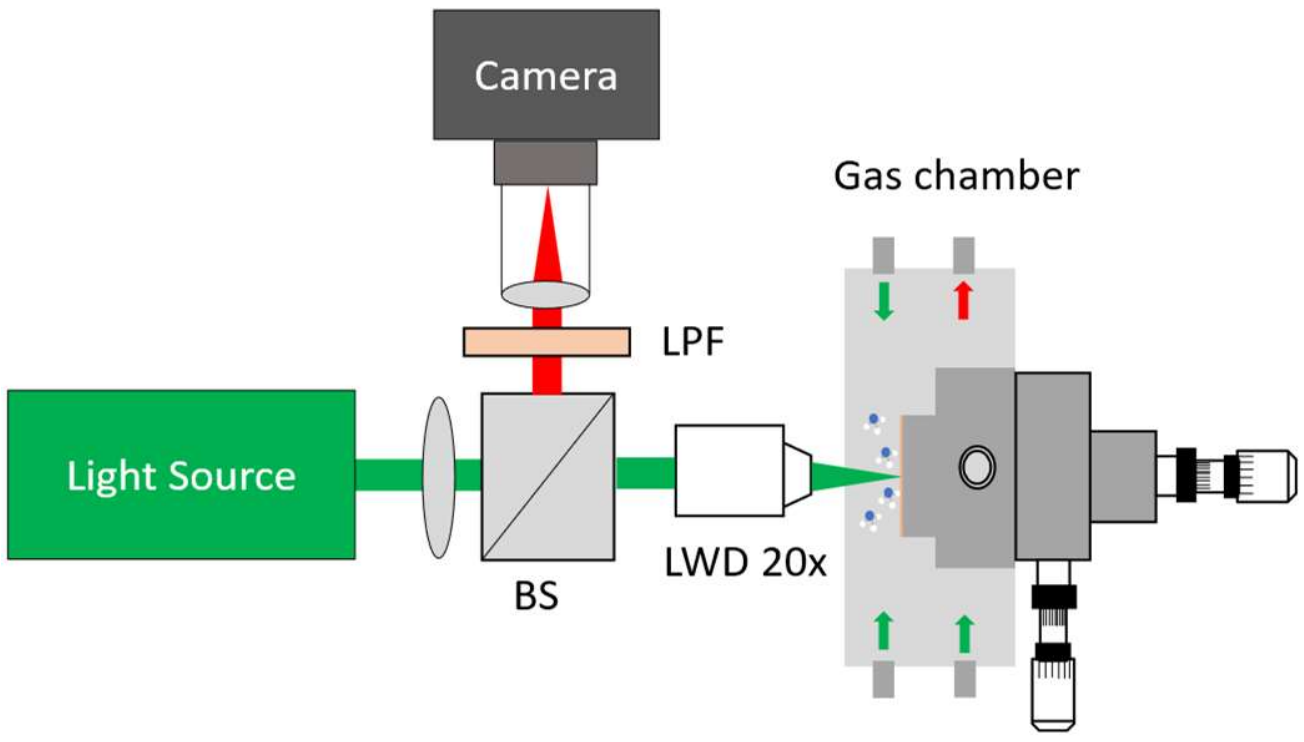

**Fig 3.** Schematic diagram of the experimental setup for gas sensing on a planar chip. LPF: long-pass filter. LWD: long-working distance objective. BS: beam splitter.

In this experiment, the PL-based sensing performance was characterized by $MoS_2$ monolayers grown on $SiO_2$/Si substrates. The PL characterization of those monolayers is discussed in Supplement S1. The Raman spectrum from one typical $MoS_2$ monolayer is shown in Supplement S2. Figure 3 illustrates the experimental measurement conducted in the reflection mode with TMD monolayers. The experimental chamber has a total volume of approximately 20 $cm^3$. A Lasos continuous-wave laser with a central wavelength of 534 nm was used to illuminate the TMD monolayers on the substrate. The PL emission was collected and filtered out by a set of long-pass filters 550 nm before the camera. Here, we focused the excitation laser on the monolayer using a long-working distance 20x objective. All target gases CO, NO, and $NO_2$ were diluted with $N_2$ to a concentration of 10 ppm and were measured in the same condition. This is not the detection limit of the $MoS_2$-based sensor but rather is merely a sample concentration. The low flow rate of the target gases (50 sccm) requires 24 seconds to flush the gas pipe, so the measurement time took 2 minutes to ensure that all the gas filled the chamber.

In Figure 4, we present the PL intensity emitted by the $MoS_2$ grown on $SiO_2$/Si in the gas chamber with the presence of target gases. Before acquiring the signal, we introduced $N_2$ for 10 minutes to set the baseline. After that, we recorded the response of $N_2$ and the target gases (CO, NO, $NO_2$) for three cycles with 2 minutes for each gas. The PL was acquired continuously during the entire process, and it was characterized as grey levels of the captured images from our CMOS camera. A small region of interest around the position where the laser beam was focused on the monolayer was selected on camera to reduce the noise from the lab environment.

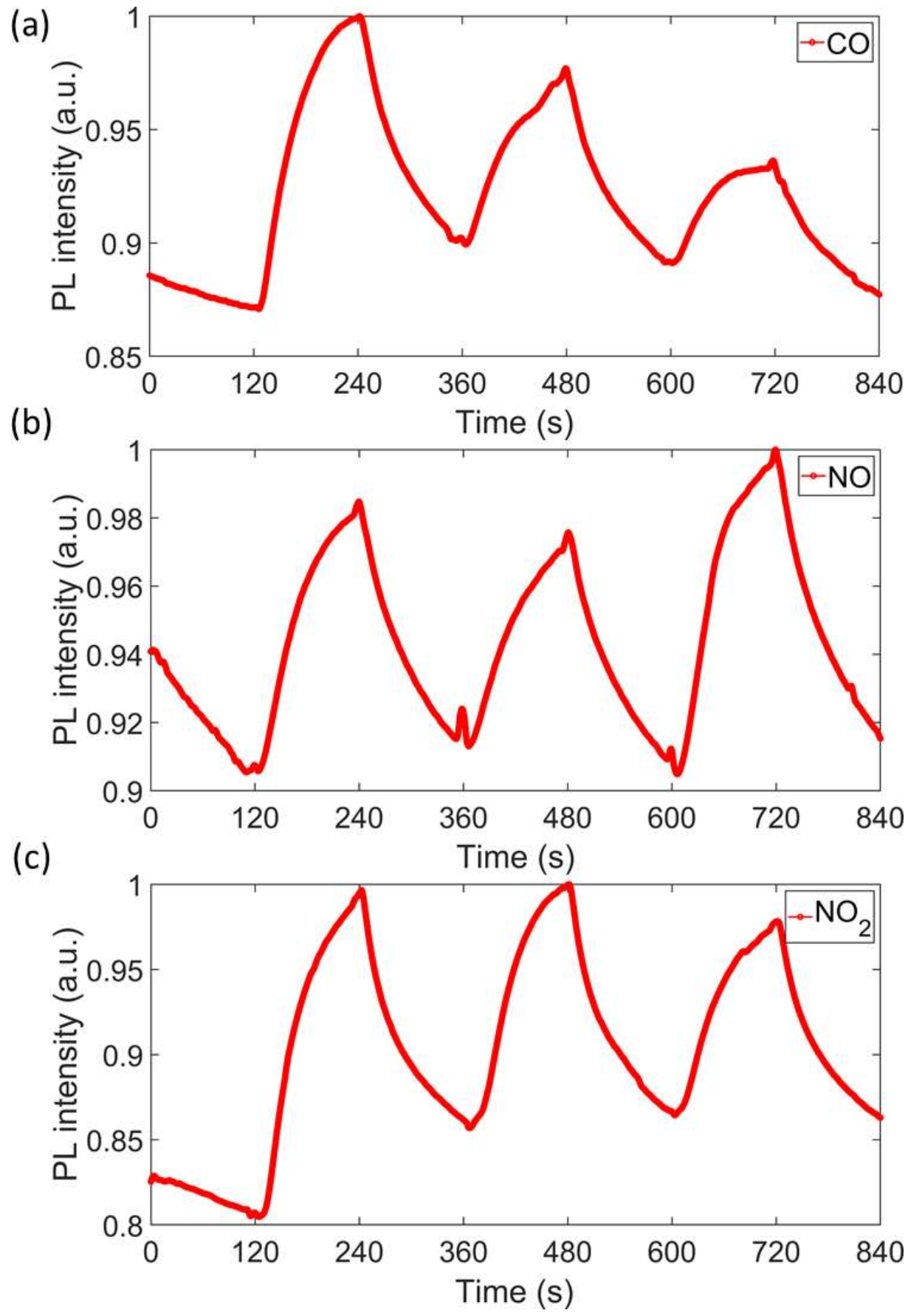

**Fig 4.** Normalized response of $MoS_2$ to the presence of (a) CO, (b) NO, (c) $NO_2$. Gas concentration is 10 ppm.

The increase of PL with CO in the chamber was observed in three cycles, and there was a drop of PL when $N_2$ was injected into the chamber to desorb CO. This finding demonstrates that CO can increase PL intensity emitted by $MoS_2$ monolayers. Because the PL intensity that we have recorded is the contribution from both excitons and trions, the increase in PL intensity is attributed to an increase in exciton intensity. The increase in PL intensity also occurred with NO and $NO_2$ in the gas chamber. All the target gases show good interaction with $MoS_2$ and modulate the PL intensity quickly, and $N_2$ can replace those gases efficiently. The response time and recovery time of $MoS_2$ in our work to the gases ranges from 60 to 120 s. This response and recovery time is significantly faster than the reported response time of 20 min and the recovery time of 30 min with $MoSe_2$ [34] or a reported response time up to 600 s [35] and up to 9 min [36] with $MoS_2$.

## 3. Conclusions

In summary, we have demonstrated optical remote sensing from TMD monolayers with the response and recovery time faster than some reported gas sensors based on field-effect transistors [34–36]. The scalable CVD-based deposition process can be expanded to other structure architectures and 2D materials, for example, fiber-based gas sensors. In addition, TMDs can also be sensitive to other environmental conditions, such as gamma rays [37], which broadens the applicability of the presented sensor scheme. The stability of photoluminescence from monolayer TMDs makes them potential photoemitters in optical gas sensing applications at room temperature. While the demonstration was carried out in a non-optimized gas chamber and by non-selective gases, we argue that a further improvement of response time and sensitivity may open great perspectives for enhanced optical gas sensing and surface-sensitive bio-analytics. All these findings open new possibilities for scalable, simple, compact lab-on-a-chip technology and waveguide-based sensing networks, which can be used for public health monitoring and environment control.

## 4. Experimental Methods

*CVD growth of TMDs:* $MoS_2$ monolayers were grown by a modified CVD technique, where a Knudsen-type effusion cell was used to deliver the sulfur precursor. Metal oxide powders such as $MoO_3$ were used as the source of transition metal atoms. Sulfur was used as the source of chalcogen atoms. Details of the method are given in Refs. [25,26]

*Density Functional Theory:* To investigate the electronic band structures and calculate the gas adsorption energy as well as charge transfer between different gases and 2D materials, we employed the Vienna ab initio Simulation Package (VASP) with a plane wave basis set [38,39]. The pseudopotentials from the projector augmented wave (PAW) were implemented [40,41]. The generalized gradient approximation of Perdew, Burke, and Ernzerhof (PBE) formalism [42] was used to account for the exchange and correlation potentials since it was proved to predict qualitatively correct results [15]. The monolayer's 5 x 5 x 1 supercell was constructed with the vacuum space at 25 Å to avoid interaction between neighboring layers. The cutoff energy of 500 eV and 5 x 5 x 1 Monkhorst-Pack mesh were selected to represent the plane-wave basis set and the Brillouin zone integration, respectively [43]. The energy convergence was set to $10^{-4}$ eV. A single gas molecule was placed on top of the monolayer at the center of the supercell. Then, all atoms were relaxed until their Hellmann-Feynman force was less than 0.02 eV/Å.

**Acknowledgements.** The Authors acknowledge the European Social Funds and the Federal State of Thuringia under Grant ID 2018FGR00088, BMBF Project SINNER Grant No.

16KIS1792, the Collaborative Research Center (CRC/SFB) 1375 NOA, the Deutsche Forschungsgemeinschaft (DFG, German Research Foundation) - Projektnummer 445275953 and under Germany's Excellence Strategy - EXC-2111-390814868, the Federal Ministry of Education and Research (BMBF) under Grant No. 13N16292, Federal Ministry for Economic Affairs and Climate Action BMWK under Grant No. 50WM2165 (QUICK3) and 50RP2200 (QuVeKS). We acknowledge support by the German Research Foundation Projekt-Nr. 512648189 and the Open Access Publication Fund of the Thueringer Universitaets- und Landesbibliothek Jena. Funding from the Development and Promotion of Science and Technology Talents Project (DPST) scholarship by the Royal Thai Government is gratefully acknowledged. The partial technical support of this research by the DFG core facility "Zentrum für Mikro- und Nanotechnologien" is grateful acknowledged. We thank Stephanie Höppener and Ulrich S. Schubert for enabling our Raman spectroscopy studies at the JCSM.

**Disclosures.** The authors declare no conflicts of interest.

**Data availability.** Data underlying the results presented in this paper are not publicly available at this time but may be obtained from the authors upon reasonable request.

**Supplemental document.** See Supplement S1−S2 for supporting content.

**References**

1. J. Zhang, X. Liu, G. Neri, *et al.*, "Nanostructured Materials for Room-Temperature Gas Sensors," Adv. Mater. **28**(5), 795–831 (2016).

2. F. Schedin, A. K. Geim, S. V. Morozov, *et al.*, "Detection of individual gas molecules adsorbed on graphene," Nat. Mater. **6**(9), 652–655 (2007).

3. K. F. Mak, C. Lee, J. Hone, *et al.*, "Atomically thin $MoS_2$: a new direct-gap semiconductor," Phys. Rev. Lett. **105**(13), 136805 (2010).

4. D. Tyagi, H. Wang, W. Huang, *et al.*, "Recent advances in two-dimensional-material-based sensing technology toward health and environmental monitoring applications," Nanoscale **12**(6), 3535–3559 (2020).

5. X. Liu, T. Ma, N. Pinna, *et al.*, "Two-Dimensional Nanostructured Materials for Gas Sensing," Adv. Funct. Mater. **27**(37), 1702168 (2017).

6. H. Li, Z. Yin, Q. He, *et al.*, "Layered Nanomaterials: Fabrication of Single- and Multilayer $MoS_2$ Film-Based Field-Effect Transistors for Sensing NO at Room Temperature," Small **8**(1), 63–67 (2012).

7. S. Yang, C. Jiang, and S. Wei, "Gas sensing in 2D materials," Appl. Phys. Rev. **4**(2), 021304 (2017).

8. M. Donarelli and L. Ottaviano, "2D materials for gas sensing applications: A review on graphene oxide, $MoS_2$, $WS_2$ and phosphorene," Sensors **18**(11), 3638 (2018).

9. B. Cho, M. G. Hahm, M. Choi, *et al.*, "Charge-transfer-based gas sensing using atomic-layer $MoS_2$," Sci. Rep. **5**, 8052 (2015).

10. F. Jiang, W. S. Zhao, and J. Zhang, "Mini-review: Recent progress in the development of $MoSe_2$ based chemical sensors and biosensors," Microelectron. Eng. **225**, 111279 (2020).

11. C. Anichini, W. Czepa, D. Pakulski, *et al.*, "Chemical sensing with 2D materials," Chem. Soc. Rev. **47**(13), 4860–4908 (2018).

12. M. Nayeri, M. Moradinasab, and M. Fathipour, "The transport and optical sensing properties of $MoS_2$, $MoSe_2$, $WS_2$ and $WSe_2$ semiconducting transition metal dichalcogenides," Semicond. Sci. Technol. **33**(2), 025002 (2018).

13. B. Huang, Z. Li, Z. Liu, *et al.*, "Adsorption of gas molecules on graphene nanoribbons and its implication for nanoscale molecule sensor," J. Phys. Chem. C **112**(35), 13442–13446 (2008).

14. S. Tongay, J. Zhou, C. Ataca, *et al.*, "Broad-range modulation of light emission in two-dimensional semiconductors by molecular physisorption gating," Nano Lett. **13**(6), 2831–2836 (2013).

15. S. Zhao, J. Xue, and W. Kang, "Gas adsorption on $MoS_2$ monolayer from first-principles calculations," Chem. Phys. Lett. **595**–**596**, 35–42 (2014).

16. Q. Yue, Z. Shao, S. Chang, *et al.*, "Adsorption of gas molecules on monolayer $MoS_2$ and effect of applied electric field," Nanoscale Res. Lett. **8**(1), 1–7 (2013).

17. J. Tian, G. Zhao, A. J. Fleisher, W. Ma, *et al.*, "Optical feedback linear cavity enhanced absorption spectroscopy," Opt. Express **29**(17), 26831–26840 (2021).

18. C. Li, K. Chen, J. Zhao, *et al.*, "High-sensitivity dynamic analysis of dissolved gas in oil based on differential photoacoustic cell," Opt. Lasers Eng. **161**, 107394 (2023).

19. Q. Wang, Z. Wang, H. Zhang, *et al.*, "Dual-comb photothermal spectroscopy," Nat. Commun. **13**(1), 2181 (2022).

20. W. Kong, F. Wan, Y. Lei, *et al.*, "Dynamic detection of decomposition gases in eco-friendly $C_5F_{10}O$ gas-insulated power equipment by fiber-enhanced Raman spectroscopy," Analytical Chemistry **96**(38), 15313–15321 (2024).

21. K. S. Pasupuleti, T. M. T. Pham, B. M. Abraham, *et al.*, "Room temperature ultrasensitive ppb-level $H_2S$ SAW gas sensor based on hybrid CuO@ $V_2C$ MXene van der Waals heterostructure," Adv. Compos. Hybrid Mater. **8***,* 132 (2025).

22. T. M. T. Pham, K. S. Pasupuleti, T. H. Jeong, *et al.*, "Facile engineering of $TiO_2$@ Pd based hybrid heterogeneous nanostructures for enhanced $NO_2$ gas sensing at room temperature under UV activation," Nanoscale **16**(48), 22252–22266 (2024).

23. G. Q. Ngo, A. George, R. T. K. Schock, *et al.*, "Scalable functionalization of optical fibers using atomically thin semiconductors," Adv. Mater. **32**(47), 2003826 (2020).

24. G. Q. Ngo, E. Najafidehaghani, Z. Gan, *et al.*, "In-fibre second-harmonic generation with embedded two-dimensional materials," Nat. Photonics **16**(11), 769–776 (2022).

25. A. George, C. Neumann, D. Kaiser, *et al.*, "Controlled growth of transition metal dichalcogenide monolayers using Knudsen-type effusion cells for the precursors," J. Phys. Mater. **2**(1), 016001 (2019).

26. E. Najafidehaghani, Z. Gan, A. George, *et al.*, "1D p–n junction electronic and optoelectronic devices from transition metal dichalcogenide lateral heterostructures grown by one-pot chemical vapor deposition synthesis," Adv. Funct. Mater. **31**(27), 2101086 (2021).

27. G. Q. Ngo, S. Ritter, C. Neumann, *et al.*, "Scalable oxide encapsulation of CVD-grown $WS_2$ and $WSe_2$ for photonic applications," Opt. Mater. Express (to be published). https://doi.org/10.1364/OME.555980

28. C. Martín, K. Kostarelos, M. Prato, *et al.*, "Biocompatibility and biodegradability of 2D materials: Graphene and beyond," Chem. Commun. **55**(39), 5540–5546 (2019).

29. S. Wang, X. Yang, L. Zhou, *et al.*, "2D nanostructures beyond graphene: Preparation, biocompatibility and biodegradation behaviors," J. Mater. Chem. B **8**(15), 2974–2989 (2020).

30. G. Q. Ngo, A. George, R. T. K. Schock, *et al.*, "Integrated Photonics: Scalable Functionalization of Optical Fibers Using Atomically Thin Semiconductors," Adv. Mater. **32**(47), 2070354 (2020).

31. A. Kuppadakkath, E. Najafidehaghani, Z. Gan, *et al.*, "Direct growth of monolayer $MoS_2$ on nanostructured silicon waveguides," Nanophotonics **11**(19), 4397–4408 (2022).

32. G. Q. Ngo, "Guided waves systems with embedded two-dimensional materials," PhD Thesis (2024), https://doi.org/10.22032/dbt.59464.

33. V. Q. Bui, T. T. Pham, D. A. Le, *et al.*, "A first-principles investigation of various gas (CO, $H_2O$, NO, and $O_2$) absorptions on a $WS_2$ monolayer: Stability and electronic properties," J. Phys. Condens. Matter **27**(30), 305005 (2015).

34. J. Baek, D. Yin, N. Liu, *et al.* "A highly sensitive chemical gas detecting transistor based on highly crystalline CVD-grown $MoSe_2$ films," Nano Res. **10**(6), 1861–1871 (2017).

35. S. Ramu, T. Chandrakalavathi, G. Murali, *et al.*, "UV enhanced NO gas sensing properties of the $MoS_2$ monolayer gas sensor," Mater. Res. Express **6**(8), 085075 (2019).

36. B. Liu, L. Chen, G. Liu, *et al.*, "High-performance chemical sensing using Schottky-contacted chemical vapor deposition grown monolayer $MoS_2$ transistors," ACS Nano **8**(5), 5304–5314 (2014).

37. T. Vogl, K. Sripathy, A. Sharma, *et al.*, "Radiation tolerance of two-dimensional material-based devices for space applications," Nat. Commun. **10**(1), 1202 (2019).

38. G. Kresse and J. Furthmüller, "Efficiency of ab-initio total energy calculations for metals and semiconductors using a plane-wave basis set," Comput. Mater. Sci. **6**(1), 15–50 (1996).

39. G. Kresse and J. Furthmüller, "Efficient iterative schemes for ab initio total-energy calculations using a plane-wave basis set," Phys. Rev. B - Condens. Matter Mater. Phys. **54**(16), 11169 (1996).

40. P. E. Blöchl, "Projector augmented-wave method," Phys. Rev. B **50**, 17953 (1994).

41. G. Kresse, and D. Joubert, "From ultrasoft pseudopotentials to the projector augmented-wave method," Phys. Rev. B - Condens. Matter Mater. Phys. **59**(3), 1758 (1999).

42. J. P. Perdew, K. Burke, and M. Ernzerhof, “Generalized gradient approximation made simple,” Phys. Rev. Lett. **77**(18), 3865 (1996).

43. H. J. Monkhorst, and J. D. Pack, “Special points for Brillouin-zone integrations,” Phys. Rev. B **13**(12), 5188 (1976).

# Photoluminescence-Based Gas Sensing with $MoS_2$ Monolayers: Supplemental information

**GIA QUYET NGO,[1,2]* CHANAPROM CHOLSUK,[1,3,4] SEBASTIAN THIELE,[5] ZIYANG GAN,[6] ANTONY GEORGE,[6] JÖRG PEZOLDT,[5] ANDREY TURCHANIN,[6] TOBIAS VOGL,[1,3,4], AND FALK EILENBERGER[1,2,7]**

[1]*Institute of Applied Physics, Abbe Center of Photonics, Friedrich Schiller University Jena, Albert-Einstein-Str. 15, 07745 Jena, Germany*

[2]*Fraunhofer-Institute for Applied Optics and Precision Engineering IOF, Albert-Einstein-Str. 7, 07745 Jena, Germany*

[3]*Department of Computer Engineering, TUM School of Computation, Information and Technology, Technical University of Munich, Theresienstr. 90, 80333 Munich, Germany*

[4]*Munich Center for Quantum Science and Technology (MCQST), Schellingstr. 4, 80799 Munich, Germany*

[5]*FG Nanotechnologie, Institute of Micro- and Nanoelectronics and Institute of Micro- and Nanotechnologies MacroNano and Institute of Materials Science and Engineering, Technical University Ilmenau, 98693 Ilmenau, Germany*

[6]*Institute of Physical Chemistry, Abbe Center of Photonics, Friedrich Schiller University Jena, Lessingstraße 10, 07743 Jena, Germany*

[7]*Max Planck School of Photonics, Germany*

**quyet.ngo@uni-jena.de*

Supplement S1 **Photoluminescence from monolayer TMDs**

Before performing gas sensing measurement, the quality of our as-grown monolayer TMDs on the substrate is confirmed by PL spectra and PL map, as displayed in Figure S1. The monolayer was grown by a CVD process. Photoluminescence mapping was carried out with a commercial confocal PL lifetime microscope (Picoquant Microtime 200), with an excitation laser operating at 530 nm. The maps were created by moving the focus of the microscope's objective over the sample, which had a magnification of 100x. The resulting spatial resolution is estimated in the range of 0.4 μm. Detection of the PL signal was carried out with a single-photon avalanche diode. Alternatively, the PL microscope was connected to a grating spectrometer (Andor Kymera 328i) equipped with a cooled CCD detector to measure PL spectra.

Figure S1a-b display the PL characterization of $MoS_2$ crystals. Here, $MoS_2$ flakes have a size ranging from 30 to 50 μm, and they exhibit A exciton at 676 nm with an FWHM of 27 nm. These characteristics are consistent with the reported value from literature data and confirm those flakes are monolayer $MoS_2$ of high quality [(1)–(4)].

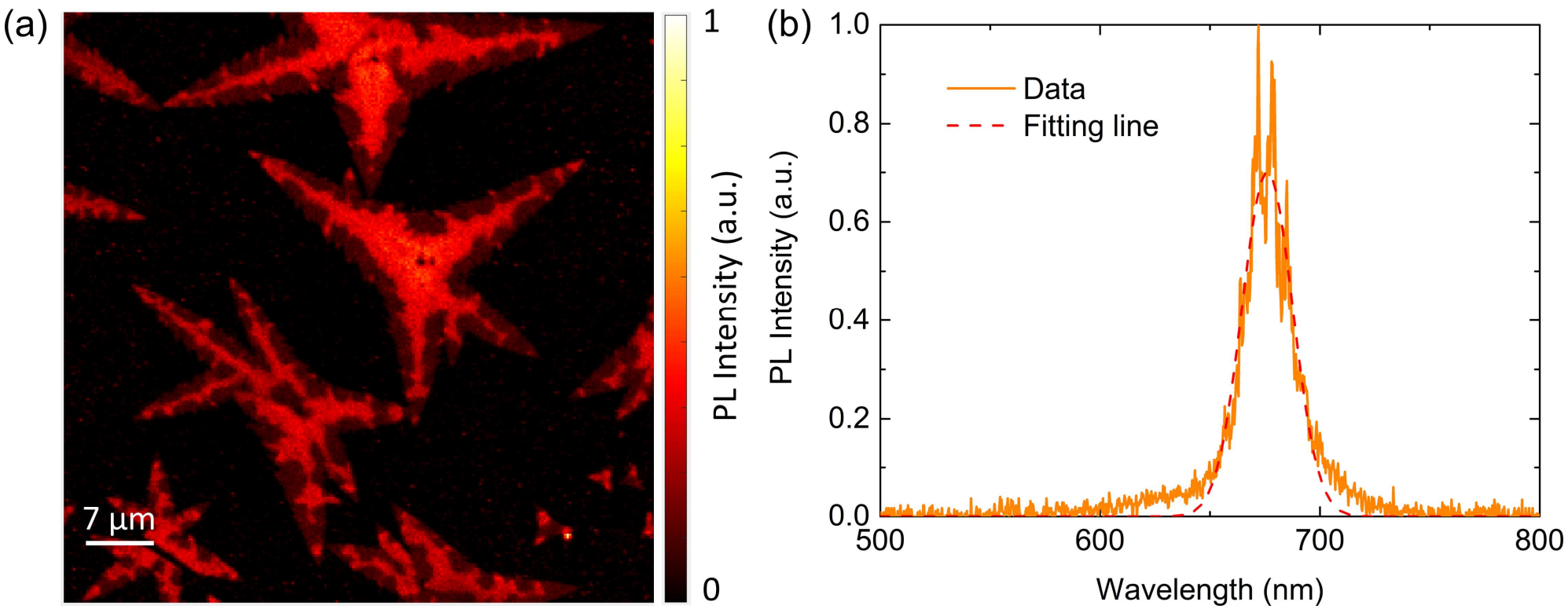

**Fig S1.** PL characterization of CVD grown monolayer TMDs before gas sensing. (a) PL map of CVD-grown $MoS_2$ monolayers. (b) Normalized PL spectrum of a monolayer in (a).

Supplement S2 **Raman spectrum from a typical $MoS_2$ monolayer**

We show the normalized Raman spectrum of one $MoS_2$ monolayer displayed in Figure S1. Raman spectroscopy was characterized using Bruker Senterra spectrometer operated in backscattering mode using 532 nm wavelength obtained with a frequency-doubled Nd:YAG Laser, a 100x objective, and a thermoelectrically cooled CCD detector. We can see a characteristic spacing of 20 $cm^{-1}$ between two Raman resonances. This spacing feature indicated the monolayer nature of the flakes [(1)–(4)].

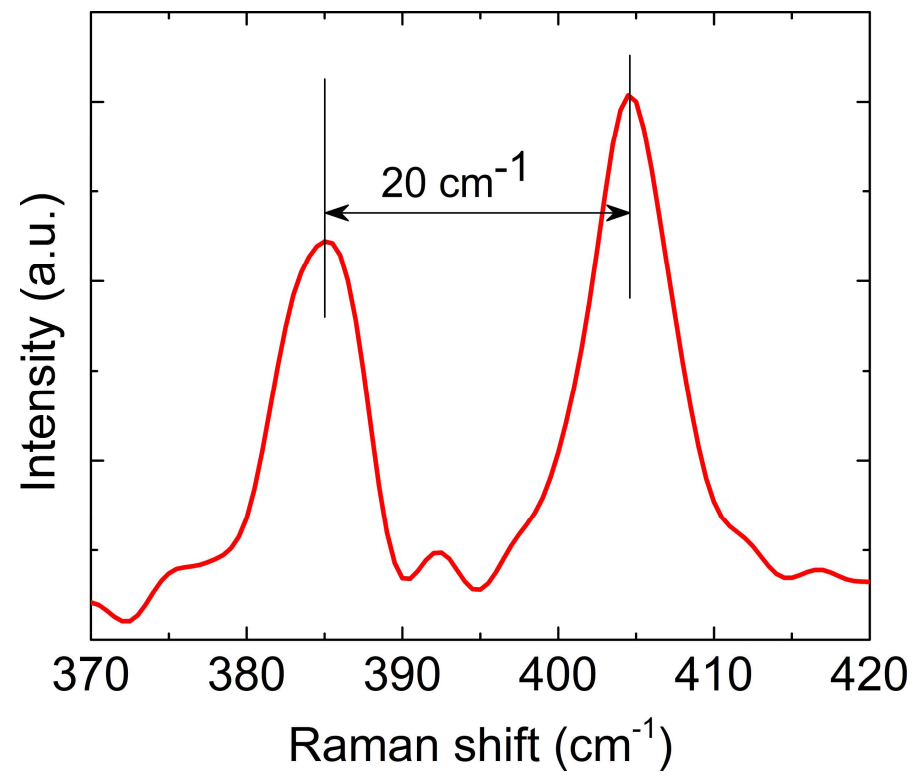

**Fig S2.** Raman characterization of CVD grown $MoS_2$ monolayer.

## References

(1) A. George, C. Neumann, D. Kaiser, *et al.*, "Controlled growth of transition metal dichalcogenide monolayers using Knudsen-type effusion cells for the precursors," J. Phys. Mater. **2**(1), 016001 (2019).

(2) G. Q. Ngo, A. George, R. T. K. Schock, *et al.*, "Scalable functionalization of optical fibers using atomically thin semiconductors," Adv. Mater. **32**(47), 2003826 (2020).

(3) G. Q. Ngo, E. Najafidehaghani, Z. Gan, *et al.*, "In-fibre second-harmonic generation with embedded two-dimensional materials," Nat. Photonics **16**(11), 769–776 (2022).

(4) A. Kuppadakkath, E. Najafidehaghani, Z. Gan, *et al.*, "Direct growth of monolayer $MoS_2$ on nanostructured silicon waveguides," Nanophotonics **11**(19), 4397–4408 (2022).